\documentclass[12pt]{article}
\usepackage{graphics}
\usepackage{graphicx}

\input{epsf}
\usepackage{cite}
\def\@fmsl@sh#1#2#3{\m@th\ooalign{$\hfil#1\mkern#2/\hfil$\crcr$#1#3$}}
 \def\eq#1\en{\begin{equation}#1\end{equation}}
\def\s[#1,#2]{[#1\stackrel{\star}{,}#2]}
\def\sx[#1,#2]{[#1\stackrel{\star_{x}}{,}#2]}

\newcommand{\nc}{\newcommand}
\nc{\beq}{\begin{equation}}
\nc{\eeq}{\end{equation}}
\nc{\beqa}{\begin{eqnarray}}
\nc{\eeqa}{\end{eqnarray}}

\def\gsim{\mathrel{\rlap{\lower4pt\hbox{\hskip1pt$\sim$}}
    \raise1pt\hbox{$>$}}}       

\begin{document}
\makeatletter
\def\fmslash{\@ifnextchar[{\fmsl@sh}{\fmsl@sh[0mu]}}
\def\fmsl@sh[#1]#2{%
  \mathchoice
    {\@fmsl@sh\displaystyle{#1}{#2}}%
    {\@fmsl@sh\textstyle{#1}{#2}}%
    {\@fmsl@sh\scriptstyle{#1}{#2}}%
    {\@fmsl@sh\scriptscriptstyle{#1}{#2}}}
\def\@fmsl@sh#1#2#3{\m@th\ooalign{$\hfil#1\mkern#2/\hfil$\crcr$#1#3$}}
\makeatother


\title{\large{\bf A Seesaw Mechanism  in the Higgs Sector}}

\author{Xavier~Calmet\thanks{xcalmet@ulb.ac.be} and Josep F. Oliver\thanks{Josep.F.Oliver@uv.es}  \\
Service de Physique Th\'eorique, CP225 \\
Boulevard du Triomphe \\
B-1050 Brussels \\
Belgium 
}

\date{June, 2006}

\maketitle

\begin{abstract}
In this letter we revisit the seesaw Higgs mechanism. We show how a seesaw mechanism in a two Higgs doublets model can trigger the electroweak symmetry breaking if at least one of the eigenvalues of the squared mass matrix is negative. We then consider two special cases of interest. In the decoupling scenario,  there is only one scalar degree of freedom in the low energy regime. In the degenerate scenario, all five degrees of freedom are in the low energy regime and will lead to observables effects at the LHC. Furthermore, in that scenario, it is possible to impose a discrete symmetry between the doublets that makes the extra neutral degrees of freedom stable. These are thus viable dark matter candidates. We find an interesting relation between the electroweak symmetry breaking mechanism and dark matter.
\end{abstract}


\newpage


During almost three decades, the key motivation for physics beyond the electroweak standard model has been to try to understand the hierarchy and the naturalness problems of the standard model (see e.g. \cite{Susskind:1978ms}). In other words why is the Higgs's boson mass so small in a first place in comparison to let us say the Planck scale and why does it remain small despite potentially large radiative corrections? Many models have been proposed to address these two problems, and most of them have now been ruled out. With the discovery of the landscape  \cite{Bousso:2000xa,Kachru:2003aw,Douglas:2003um,Susskind:2003kw,Denef:2004ze}  in string theory, the guidance principles for model building  have evolved. The fine tuning problems of the standard models have been put aside and new phenomenological models have been proposed (see e.g. \cite{Arkani-Hamed:2004fb,Giudice:2004tc,Calmet:2004ck,Manohar:2006gz}). But, from our perspective,  even if nature is really fine-tuned, it remains the issue  of explaining  why the electroweak gauge symmetry is broken. There is always a price to break a symmetry  and we can hope that triggering the spontaneous symmetry breaking requires a mechanism that could be unveiled at the Tevatron or the LHC. In other words,  the mechanism that renders the Higgs's boson mass imaginary will hopefully be observable at the Tevatron or the LHC otherwise it will be very complicated to gain some insight into the physics beyond the standard model.

A lot of effort has recently been invested in simple extensions of the standard model which involve scalar singlets \cite{vanderBij:2006ne,vanderBij:2006pg,Hill:1987ea,Patt:2006fw}. The simple observation in these papers is that the Higgs's boson mass term in the standard model action 
\begin{eqnarray} \label{ophmass}
m^2 \phi^\dagger \phi
\end{eqnarray}
is the only operator  which is super-renormalizable and one could imagine coupling singlets in a renormalizable way to the standard model. For example, one could consider coupling the standard model to a technicolor sector through that operator
\begin{eqnarray}
\frac{<\bar f f>}{\Lambda_{TC}} \phi^\dagger \phi
\end{eqnarray}
where $f$ are new strongly interacting fermions which transform under the technicolor group. In that framework the technicolor sector would trigger the Higgs mechanism and a negative mass term does not  need to be introduced by hand. It has been shown in \cite{Calmet:2003uj} that the operator (\ref{ophmass}) can open the door to a hidden sector with an energy scale in the weak scale range.

Here we shall consider another option and include one more scalar SU(2) doublet in the Higgs sector and consider different scenarios for the mass texture of this two Higgs doublets model. We consider a low energy theory and assume that the mass texture we obtain for the scalar doublets is the result of some more fundamental interaction with an energy scale which is way above the weak scale.
The nice feature of these models is that they generate a negative squared mass for of the Higgs doublet thereby triggering the Higgs mechanism.

Besides discovering at least one Higgs boson, LHC could allow to produce and detect dark matter. There are strong astrophysical evidences for dark matter coming from different observations  (see e.g. the review on dark matter in the Review of particle physics \cite{Eidelman:2004wy}), but there is, to date, no evidence coming from laboratory experiments.  In this letter we will describe a scenario to generate the Higgs mechanism.  There are two special limits of this model.  In the decoupling scenario,  there is only one scalar degree of freedom in the low energy regime. In the degenerate scenario, all five degrees of freedom are in the low energy regime and will lead to observables effects at the LHC. Furthermore, in that scenario, it is possible to impose a discrete symmetry between the doublets that makes the extra neutral degrees of freedom stable. If this scenario is correct, the LHC will not only unveil the mechanism for spontaneous symmetry breaking, but it will also produce dark matter.

We shall start with a generic two scalar doublets model. The action of the scalar potential sector is given by 
\begin{eqnarray} \label{2HD}
S_{scalar}&=& - \int d^4x \left( h_a^\dagger  h_b^\dagger \right )
\left(\begin{array}{cc} m^2_a & m_c^2 \\ m_c^2 & m_b^2 \end{array}
\right )
\left(\begin{array}{c} h_a  \\ h_b \end{array}
\right ) - \\ \nonumber &&  
- \int d^4x \left (
   \frac{\lambda_a}{2} (h_a^\dagger h_a)^2
+  \frac{\lambda_b}{2}  (h_b^\dagger h_b)^2
+  \lambda_c (h_a^\dagger h_b)(h_a^\dagger h_b) \right.
\\ \nonumber && \left .
+  \frac{ \lambda_d}{2}  (h_a^\dagger h_a) (h_b^\dagger h_b) 
+   \frac{\lambda_e}{2}  (h_a^\dagger h_b) (h_b^\dagger h_a)  
+ \lambda_f  (h_a^\dagger h_a) (h_a^\dagger h_b)
 +  \lambda_g  (h_b^\dagger h_b) (h_b^\dagger h_a) + h.c.
\right).
\end{eqnarray}
We assume that $m_a$ and $m_b$ are real since we do not want to trigger the gauge symmetry through an imaginary mass term put by hand in the Lagrangian. On the other hand we think of the cross-terms proportional to $m_c$ as of interactions coming from high energy physics and we include complex numbers in the parameter range  of our model. The mass matrix 
\begin{eqnarray} \label{massmatrix}
\left(\begin{array}{cc} m^2_a & m_c^2 \\ m_c^2 & m_b^2 \end{array}
\right )
\end{eqnarray}
can be diagonalized and one obtains:
\begin{eqnarray} \label{diagmat}
\left(\begin{array}{cc} m^2_u & 0 \\ 0 & m^2_d \end{array}
\right )
\end{eqnarray}
with
\begin{eqnarray} 
m_u^2&=&\frac{-m_a^4-4 m_c^4-m_b^4+2 m_b^2m_a^2 + (m_a^2+m_b^2)\sqrt{m_a^4-2 m_b^2m_a^2+4 m_c^4+m_b^4}}{2 \sqrt{m_a^4-2 m_b^2m_a^2+4 m_c^4+m_b^4}}
\end{eqnarray}
and
\begin{eqnarray} 
m_d^2&=&\frac{m_a^4+4 m_c^4+m_b^4-2 m_b^2m_a^2 + (m_a^2+m_b^2)\sqrt{m_a^4-2 m_b^2m_a^2+4 m_c^4+m_b^4}}{2 \sqrt{m_a^4-2 m_b^2m_a^2+4 m_c^4+m_b^4}}.
\end{eqnarray}
We shall require that one of these squared masses be negative in order to generate the electroweak symmetry breaking.

There are two limit cases of interest. One is the decoupling case which has been proposed in \cite{Calmet:2002rf} (see \cite{Atwood:2004mj} for an interesting extension of the idea). 
The seesaw Higgs mechanism  \cite{Calmet:2002rf} in its simplest form allows to trigger the Higgs mechanism.  We consider the same action as that of the standard model but with a
modified scalar sector.  The first scalar boson is denoted by $h$ and
the second scalar doublet by $H$.  Both doublets have exactly the same
quantum numbers as the usual standard model Higgs doublet. In a first approximation, the Yukawa
sector involves only the boson $h$.  The scalar potential is chosen (i.e. fine-tuned)
according to
\begin{eqnarray} \label{action1}
S_{scalar}&=& - \int d^4x \left( h^\dagger  H^\dagger \right )
\left(\begin{array}{cc} 0 & m^2 \\ m^2 & M^2 \end{array}
\right )
\left(\begin{array}{c} h  \\ H \end{array}
\right ) - \\ \nonumber &&  - \int d^4x \lambda_h (h^\dagger h)^2
-  \int d^4x \lambda_H
(H^\dagger H)^2
\end{eqnarray}
i.e. we assume that the first boson $h$ is almost  massless whereas the second
boson $H$ is massive after renormalization. We assume that $m$ is real. We note that this  mass texture can be generated in the framework of an invisible technicolor model \cite{Haba:2005jq} or in topcolor models \cite{Dobrescu:1997nm,Chivukula:1998wd,He:2001fz}. It  can also emerge in supersymmetric models \cite{Ito:2000cj}.  After diagonalization of the mass matrix in eq. (\ref{action1}) using
\begin{eqnarray} \label{rotation}
R=\left(\begin{array}{cc} 1& \frac{m^2}{M^2}
    \\  -\frac{m^2}{M^2} & 1 \end{array}
\right ) \approx \left(\begin{array}{cc} 1& 0
    \\ 0 & 1 \end{array}
\right ),
\end{eqnarray}
we obtain the squared masses of the mass eigenstates
\begin{eqnarray} \label{massmatrix2}
{\cal M}^2 \approx
\left(\begin{array}{cc} - \frac{m^4}{M^2} & 0 \\ 0 & M^2 \end{array}
\right ).
\end{eqnarray}
Note that the minus sign is not trivial as it is in the fermionic seesaw \cite{seesaw} case where it can be reabsorbed by a fermion phase transformation. The first boson $h$ has become a Higgs boson with a negative squared
mass given by
\begin{eqnarray}
m_h^2=- \frac{m^4}{M^2}
\end{eqnarray}
whereas the second scalar boson $H$ has a positive squared mass of the
order of the large squared scale $M^2$ and is thus not contributing to the
electroweak symmetry breaking. The mass of the physical Higgs boson is
given by
\begin{eqnarray} \label{physhiggsmass}
M^{phys}_h=\sqrt{2}\frac{m^2}{M}.
\end{eqnarray}
In that scenario one can imagine that the heavy Higgs boson decouples completely from the low energy regime. One finds that a Higgs boson mass of the order of 100 GeV can be
obtained if $m\sim 10^9$ GeV and $M\sim\Lambda_{GUT}\sim 10^{16}$ GeV. This would correspond to a decoupling scenario. Note however that the decoupling needs not to be as total and the mass scale $M$ could easily be in the TeV range. The couplings of the scalar doublet $H$, which involves one neutral scalar particle, a pseudo-scalar boson and a charged scalar boson, to the fermions are assumed to be very small.

We shall now consider another interesting case. Let us again consider the same action as that of the standard model but with a modified scalar sector.  The first scalar boson is denoted by $h_a$ and
the second scalar doublet by $h_b$.  Both doublets have exactly the same
quantum numbers as the usual standard model Higgs doublet.  The scalar potential is chosen
to be
\begin{eqnarray} \label{action2}
S_{scalar}&=& - \int d^4x \left( h_a^\dagger  h_b^\dagger \right )
\left(\begin{array}{cc} 0 & m_c^2 \\ m_c^2 & 0 \end{array}
\right )
\left(\begin{array}{c} h_a  \\ h_b \end{array}
\right ) - \\ \nonumber &&  
- \int d^4x \left (
\frac{\lambda_a}{2} (h_a^\dagger h_a)^2
+ \frac{ \lambda_b}{2}  (h_b^\dagger h_b)^2
+  \lambda_c (h_a^\dagger h_b)(h_a^\dagger h_b) 
  \right .
\\ \nonumber && \left .
+ \frac{ \lambda_d}{2}  (h_a^\dagger h_a) (h_b^\dagger h_b) 
+  \frac{\lambda_e}{2}  (h_a^\dagger h_b) (h_b^\dagger h_a) 
+  \lambda_f  (h_a^\dagger h_a) (h_a^\dagger h_b)
 +  \lambda_g  (h_b^\dagger h_b) (h_b^\dagger h_a) + h.c.
\right),
\end{eqnarray}
i.e. we set $m_a^2=m^2_b=0$ in the action given in (\ref{2HD}).  The Yukawa sector is given by
\begin{eqnarray}
S_{Y} =-\int d^4x \sum_{ij} (Y^{(d)}_{ij} \bar L_i h_a r_j + W^{(d)}_{ij}  \bar L_i h_b r_j) + 
\sum_{ij} (Y^{(u)}_{ij} \bar L_i \bar h_a r_j + W^{(u)}_{ij}  \bar L_i \bar h_b r_j) +h.c.
\end{eqnarray}
where $\bar h_i=i \sigma_2 h^*_i$, $L_i$ are the relevant left-handed fermionic fields and $r_i$ the relevant right-handed ones.

We can now diagonalize the mass matrix and find:
\begin{eqnarray} \label{massmatrix3}
\left(\begin{array}{cc} 0 &m_c^2 \\ m_c^2 & 0 \end{array}
\right ) \to 
\left(\begin{array}{cc} m_c^2 & 0 \\ 0& - m_c^2  \end{array}
\right ).
\end{eqnarray}
i.e.  the two doublets are still degenerate in mass but one of them has a negative squared mass and thus acquires a vacuum expectation  value.  The mass eigenstates are given by
\begin{eqnarray} 
\left(\begin{array}{c}  h \\  H \end{array}
\right )= \frac{1}{\sqrt{2}}
\left(\begin{array}{cc} 1 &1 \\-1 & 1 \end{array} \right)
\left(\begin{array}{c}  h_a \\  h_b \end{array}
\right )
\end{eqnarray}
Let us now study the Yukawa sector of this model. One option would be to decouple the doublet that does not contribute to the electroweak symmetry breaking by fine-tuning the Yukawa couplings of that scalar doublet to the standard model fermion. This would suppress the neutral flavor current transitions that appear at tree level and are thus dangerous. It is however possible and more natural to  impose a discrete symmetry between the two doublets $h_a\to h_b$ and $h_b\to h_a$ and the Yukawa couplings involving the doublet $h_b$ are thus diagonalized at the same time as those involving the doublet $h_a$ and which gives rise to the fermion masses.  To insure that the scalar doublet which breaks the gauge symmetry spontaneously is the one that couples to fermions, we choose $m_c^2=-m^2$. In terms of the mass eigenstates we find:
\begin{eqnarray}
S_Y =-\int d^4x \sum_{ij} Y^{(d)}_{ij} \bar L_i h r_j  + 
\sum_{ij} W^{(u)}_{ij} \bar L_i \bar h r_j +h.c.
\end{eqnarray}
i.e. only the scalar doublet which generates the symmetry breaking couples to the fermions and hereby gives them a mass. In other words we choose to give a vacuum expectation value to the scalar doublet which has positive parity under the symmetry  $h_a\to h_b$ and $h_b\to h_a$. The discrete symmetry between $h_a$ and $h_b$ implies relations between the parameters of the action:  $\lambda_a=\lambda_b=\lambda_1$,  $\lambda_c=\lambda_c^*=\lambda_5$, $\lambda_d=\lambda_2$, $\lambda_e=\lambda_3$ and $\lambda_f=\lambda_g=\lambda_4$. We then have 
\begin{eqnarray} 
S_{scalar}&=& - \int d^4x \left( h_a^\dagger  h_b^\dagger \right )
\left(\begin{array}{cc} 0 & -m^2 \\- m^2 & 0 \end{array}
\right )
\left(\begin{array}{c} h_a  \\ h_b \end{array}
\right ) - \\ \nonumber &&  
- \int d^4x \left (
   \lambda_1 \left [ (h_a^\dagger h_a)^2+  (h_b^\dagger h_b)^2 \right]
+  \lambda_2 (h_a^\dagger h_a)(h_b^\dagger h_b)
 + \lambda_3 (h_a^\dagger h_b)(h_b^\dagger h_a)  \right .
\\ \nonumber && \left .
+  \left (\lambda_4 \left[  (h_a^\dagger h_a) (h_a^\dagger h_b) +  (h_b^\dagger h_b) (h_b^\dagger h_a) \right] +h.c. \right )
+  \lambda_5  \left[   (h_a^\dagger h_b) (h_a^\dagger h_b) +  (h_b^\dagger h_a) (h_b^\dagger h_a) \right]
\right).
\end{eqnarray}
In terms of the mass eigenstates $h$ and $H$ which in the unitary gauge are given by
\begin{eqnarray} 
h = \left(\begin{array}{c} 0 \\ \frac{\phi+v}{\sqrt{2}} \end{array}
\right ) \ \mbox{and} \  H =  \left(\begin{array}{c} h^+ \\ \frac{\eta+i A^0}{\sqrt{2}} \end{array}
\right )
\end{eqnarray}
 the scalar part of the action reads
\begin{eqnarray} \label{newaction}
S_{scalar}&=&  \int d^4x \left( m^2 h^\dagger h -m^2 H^\dagger H + \lambda_h (h^\dagger h)^2
+ \lambda_H (H^\dagger H)^2  \right . \\ \nonumber && \left .
+\rho_1 (H^\dagger H) (h^\dagger h) + \rho_2 (H^\dagger h) (h^\dagger H)
+\left (\rho_3 (H^\dagger h)^2+ h.c. \right) 
\right),
\end{eqnarray}
with 
\begin{eqnarray} 
\lambda_h&=& \frac{1}{4} ( 2 \lambda_1 +\lambda_2 +\lambda_3 + 4 {\mbox Re} \lambda_4 + 2 \lambda_5) \\ \nonumber
\lambda_H&=& \frac{1}{4} ( 2 \lambda_1 +\lambda_2 +\lambda_3 - 4 {\mbox Re} \lambda_4 + 2 \lambda_5) \\ \nonumber
\rho_1&=& \frac{1}{2} (2\lambda_1+\lambda_2 -\lambda_3 - 2\lambda_5) \\ \nonumber
\rho_2&=& \frac{1}{2} (2\lambda_1-\lambda_2 +\lambda_3 - 2\lambda_5) \\ \nonumber
\rho_3&=&\frac{1}{4} ( 2 \lambda_1 - \lambda_2 - \lambda_3 + 4 i  {\mbox Im}  \lambda_4 + 2 \lambda_5),
\end{eqnarray}
note that $\rho_3$ and $\lambda_4$ can be made real by performing a phase redefinition of $H$.

It is easy to see that $h$ acquires a vacuum expectation value given by $v^2=2 m^2/\lambda_h$ whereas $H$ does not contribute to the electroweak symmetry breaking.
Note that the symmetry $h_{a/b}\to h_{b/a}$ can now be interpreted as a $Z_2$ symmetry under which 
$h$ has even parity and $H$ has odd parity. The mass spectrum of the scalar sector of our model is given by:
\begin{eqnarray} 
m_\phi^2&=&2 \lambda_h v^2 \\ 
m_\eta^2&=&2 m^2+(\rho_1+\rho_2+\rho_3) v^2\\
m_{A^0}^2&=&2 m^2+(\rho_1+\rho_2-\rho_3) v^2\\
m_{h^\pm}^2&=& m^2+ \rho_1 v^2.
\end{eqnarray}

The phenomenology of the degenerate seesaw mechanism is quite interesting. Note that our model is similar to a type I two Higgs doublets model, however, in our case only one of the scalars develops a vacuum expectation value. We have two neutral scalar fields $\phi$ and $\eta$,  a pseudo-scalar $A^0$ and two charged scalars $h^\pm$. The scalar $\phi$ that triggers the electroweak symmetry breaking can couple to two gauge bosons e.g. $\phi \to Z Z$ or $\phi \to W^+ W^-$, however, the coupling of $\eta$ to the gauge bosons always involve two scalar particles i.e. $\eta + \eta \to Z Z $, $\eta + \eta \to W^+ W^-$ or $Z \to \eta+\eta$. This is a consequence of the $Z_2$ symmetry. Thus at LEP they could only have been produced in pairs and therefore only the lightest first neutral scalar Higgs boson $\phi$ could have been produced provided its mass is below 114 GeV. The LEP production mechanism for our Higgs scalar is as in the standard model and therefore it can be produced alone. Furthermore, at the LHC,  the main production channel for a fairly light scalar Higgs boson $\phi$  of the order of 120 GeV is through gluons fusion and involves the Yukawa coupling to the top quark. Since only $\phi$ couples to the fermions, it will be the only scalar produced through this reaction
\begin{eqnarray}
\Gamma_{Seesaw}(GG\to  \phi)= \Gamma_{SM}(GG\to \phi).
\end{eqnarray}
We note that unitarity of the S matrix is restored by the boson $\phi$.
It will however  be possible to differentiate this model from the standard model by observing obviously the charged scalar bosons and the pseudo-scalar boson but also by studying the Drell-Yan production, where both neutral scalars $\phi$ and $\eta$ can  be produced in pairs. 

It is interesting to have an estimate of the production cross-section of these new particles.  For the sake of simplicity we assume $\rho_1=\rho_2=\rho_3=0$, i.e. the particles of the second doublet are degenerate in mass.  This simplification enable us to treat the neutral scalars $\eta$ and $A^0$ as a single (complex) field $h^0=(\eta+i A^0)/\sqrt{2}$. This scalar will be produced via a Drell-Yan process.

\begin{figure}
\includegraphics[scale=0.4,angle=0]{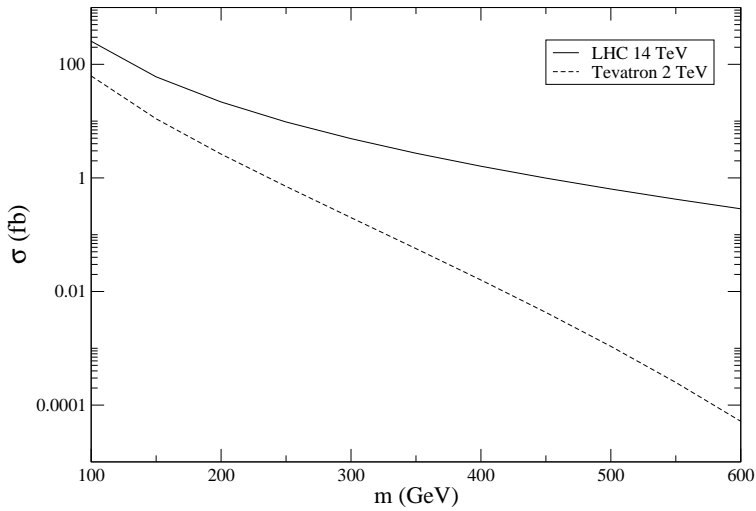}
\caption{Pair production cross-sections of $h^0$. We have used CompHEP to obtain the leading order correction and applied a K-factor of $1.25$ for the LHC and $1.3$ for the Tevatron.}
\label{ppxsecs}
\end{figure}

The $h^0$ pair production cross-section at the next to leading order for LHC and Tevatron is displayed in figure (\ref{ppxsecs}).
 We use CompHEP \cite{Pukhov:1999gg}  to compute the cross-sections to leading order and then we include a K-factor of $1.25$ for the LHC and $1.3$ for the Tevatron to take into account the next to leading order corrections, see ref. \cite{Muhlleitner:2003me}.

Note that  the Drell-Yan channel is not available for a standard model like Higgs boson. In this case the important channels are gluon-gluon fusion as well as W-bosons fusion (see e.g. \cite{Djouadi:2005gi}).  Using CompHEP we have made a  rough estimate  of the production  channels involving W-bosons or Z-bosons fusions  and found that they are subdominant with respect to the the Drell-Yan production.

We would like to finally point out that this kind of model has received renewed interest over the last few years. The addition to the standard model  of extra degrees of freedom in the Higgs boson sector  can affect the quadratic divergences to the Higgs boson squared mass. It is possible that new physics will cancel, totally or partially, these quadratic corrections, therefore allowing to shift the naturalness cutoff from 1 TeV to a higher scale \cite{Calmet:2003uj,Barbieri:2005kf,Barbieri:2006dq}. 
Another consequence of adding extra degrees of freedom to the standard model is that the contributions to electroweak precision observables are modified, possibly allowing for a heavier Higgs boson of a mass of about 500 GeV.  In ref. \cite{Ma:2006fn} a similar model has been investigated and it has been proposed that neutrino masses, baryogenesis and dark matter could have a common origin.

One of the interesting new features of our model is the connection between the electroweak symmetry breaking mechanism and dark matter which, if our mechanism is the one chosen by nature, would imply that dark matter will be produced at the LHC. Indeed  the $Z_2$ symmetry under which the particles  $\eta$, $A^0$ and $h^\pm$ are odd,   implies that the lightest of them is completely stable. We therefore have a natural dark matter candidate in our model.  In that sense, this model is linking the spontaneous symmetry breaking mechanism to dark matter and implies that the dark matter particle has a mass comparable to that of the Higgs boson or at least that some components of the dark matter are linked to an extended Higgs sector.

\subsection*{Acknowledgments}
\noindent 
X.C. would like to thank Elizabeth Jenkins for a stimulating discussion during his visit at UC San Diego.  This work was supported in part by the IISN and the Belgian science
policy office (IAP V/27).

\bigskip



\baselineskip=1.6pt
\begin{thebibliography}{99}

\bibitem{Susskind:1978ms}
  L.~Susskind,
  Phys.\ Rev.\ D {\bf 20}, 2619 (1979).
  
\bibitem{Bousso:2000xa}
  R.~Bousso and J.~Polchinski,
  JHEP {\bf 0006}, 006 (2000)
  [arXiv:hep-th/0004134].
\bibitem{Kachru:2003aw}
  S.~Kachru, R.~Kallosh, A.~Linde and S.~P.~Trivedi,
  Phys.\ Rev.\ D {\bf 68}, 046005 (2003)
  [arXiv:hep-th/0301240].
\bibitem{Douglas:2003um}
  M.~R.~Douglas,
  JHEP {\bf 0305}, 046 (2003)
  [arXiv:hep-th/0303194].

\bibitem{Susskind:2003kw}
  L.~Susskind,
  arXiv:hep-th/0302219.
  
\bibitem{Denef:2004ze}
  F.~Denef and M.~R.~Douglas,
  JHEP {\bf 0405}, 072 (2004)
  [arXiv:hep-th/0404116].
  
   
\bibitem{Arkani-Hamed:2004fb}
  N.~Arkani-Hamed and S.~Dimopoulos,
  JHEP {\bf 0506}, 073 (2005)
  [arXiv:hep-th/0405159].
  
\bibitem{Giudice:2004tc}
  G.~F.~Giudice and A.~Romanino,
  Nucl.\ Phys.\ B {\bf 699}, 65 (2004)
  [Erratum-ibid.\ B {\bf 706}, 65 (2005)]
  [arXiv:hep-ph/0406088].

\bibitem{Calmet:2004ck}
  X.~Calmet,
  Eur.\ Phys.\ J.\ C {\bf 41}, 245 (2005)
  [arXiv:hep-ph/0406314].
  
\bibitem{Manohar:2006gz}
  A.~V.~Manohar and M.~B.~Wise,
  Phys.\ Lett.\ B {\bf 636}, 107 (2006)
  [arXiv:hep-ph/0601212].
  
\bibitem{vanderBij:2006ne}
  J.~J.~van der Bij,
  Phys.\ Lett.\ B {\bf 636}, 56 (2006)
  [arXiv:hep-ph/0603082].
\bibitem{vanderBij:2006pg}
  J.~J.~van der Bij and S.~Dilcher,
  arXiv:hep-ph/0605008.
\bibitem{Hill:1987ea}
  A.~Hill and J.~J.~van der Bij,
  Phys.\ Rev.\ D {\bf 36}, 3463 (1987).
  
\bibitem{Patt:2006fw}
  B.~Patt and F.~Wilczek,
  arXiv:hep-ph/0605188.
  

\bibitem{Calmet:2003uj}
  X.~Calmet,
  Eur.\ Phys.\ J.\ C {\bf 32}, 121 (2003)
  [arXiv:hep-ph/0302056].
  
  
  
\bibitem{Eidelman:2004wy}
  S.~Eidelman {\it et al.}  [Particle Data Group],
  Phys.\ Lett.\ B {\bf 592}, 1 (2004).

\bibitem{Calmet:2002rf}
  X.~Calmet,
  Eur.\ Phys.\ J.\ C {\bf 28}, 451 (2003)
  [arXiv:hep-ph/0206091].
  
\bibitem{Atwood:2004mj}
  D.~Atwood, S.~Bar-Shalom and A.~Soni,
  Eur.\ Phys.\ J.\ C {\bf 45}, 219 (2006)
  [arXiv:hep-ph/0408191];
  S.~Bar-Shalom, D.~Atwood and A.~Soni,
  PoS {\bf HEP2005}, 358 (2006)
  [arXiv:hep-ph/0511281].

\bibitem{Haba:2005jq}
  N.~Haba, N.~Kitazawa and N.~Okada,
  arXiv:hep-ph/0504279.

\bibitem{Dobrescu:1997nm}
B.~A.~Dobrescu and C.~T.~Hill,
Phys.\ Rev.\ Lett.\  {\bf 81}, 2634 (1998)
[arXiv:hep-ph/9712319];

\bibitem{Chivukula:1998wd}
R.~S.~Chivukula, B.~A.~Dobrescu, H.~Georgi and C.~T.~Hill,
Phys.\ Rev.\ D {\bf 59}, 075003 (1999)
[arXiv:hep-ph/9809470];

\bibitem{He:2001fz}
H.~J.~He, C.~T.~Hill and T.~M.~Tait,
Phys.\ Rev.\ D {\bf 65}, 055006 (2002)
[arXiv:hep-ph/0108041].
\bibitem{Ito:2000cj}
M.~Ito,
Prog.\ Theor.\ Phys.\  {\bf 106}, 577 (2001)
[arXiv:hep-ph/0011004].

\bibitem{seesaw}
  P.~Minkowski,
  Phys.\ Lett.\ B {\bf 67}, 421 (1977);
  T. Yanagida,
  in {\it Proceedings of the Workshop on the Baryon Number of the 
  Universe and Unified Theories}, Tsukuba, Japan, 13-14 Feb 1979;
  M. Gell-Mann, P. Ramond, and R. Slansky,
  in {\it Supergravity}, North Holland, Amsterdam, 1979;
  S.~Glashow,
  NATO\ Adv.\ Study\ Inst.\ Ser.\ B\ Phys.\ {\bf 59}, 687 (1979);
R.~N.~Mohapatra and G.~Senjanovic, Phys. Rev. Lett. {\bf 44}, 912 (1980).



\bibitem{Pukhov:1999gg}
  A.~Pukhov {\it et al.},
  arXiv:hep-ph/9908288; 
  E.~Boos {\it et al.}  [CompHEP Collaboration],
  Nucl.\ Instrum.\ Meth.\ A {\bf 534}, 250 (2004)
  [arXiv:hep-ph/0403113].
  

  
\bibitem{Muhlleitner:2003me}
  M.~Muhlleitner and M.~Spira,
  Phys.\ Rev.\ D {\bf 68}, 117701 (2003)
  [arXiv:hep-ph/0305288].
  
  
\bibitem{Djouadi:2005gi}
  A.~Djouadi,
  arXiv:hep-ph/0503172.
  
  

\bibitem{Barbieri:2005kf}
  R.~Barbieri and L.~J.~Hall,
  arXiv:hep-ph/0510243.


\bibitem{Barbieri:2006dq}
  R.~Barbieri, L.~J.~Hall and V.~S.~Rychkov,
  arXiv:hep-ph/0603188.
  
   
\bibitem{Ma:2006fn}
  E.~Ma,
  arXiv:hep-ph/0605180.
  
 
  
  
\end{thebibliography}
\end{document}